\documentclass{aa}  

\usepackage{graphicx}
\usepackage{txfonts}
\usepackage{dcolumn} %za alignment u tablici
\usepackage{hyperref}
\begin{document}

   \title{Determination of CME orientation and consequences for their propagation}

   %\subtitle{I. Overviewing the $\kappa$-mechanism}

   \author{K. Martini\'c,
          \inst{1}
          %\and
          M. Dumbovi\'c,\inst{1}
          M. Temmer, \inst{2}
          A. Veronig, \inst{2,}\inst{3}
          B. Vr\v{s}nak \inst{1}
          }

   \institute{Hvar Observatory, Faculty of Geodesy, University of Zagreb, Zagreb, Croatia\\
              \email{kmartinic@goef.hr}
         \and
             Institute of Physics, University of Graz, Graz, Austria
        \and
             Kanzelhöhe Observatory for Solar and Environmental Research, University of Graz, Austria\\
             }

   %\date{Received September 15, 1996; accepted March 16, 1997}

% \abstract{}{}{}{}{} 
% 5 {} token are mandatory
 
  \abstract
   {The configuration of the interplanetary magnetic field and features of the related ambient solar wind in the ecliptic and meridional plane are different. Therefore, one can expect that the orientation of the flux-rope axis of a coronal mass ejection (CME) influences the propagation of the CME itself. However, the determination of the CME orientation, especially from image data, remains a challenging task to perform.}
   {This study aims to provide a reference to different CME orientation determination methods in the near-Sun environment. Also, it aims to investigate the non-radial flow in the sheath region of the interplanetary CME (ICME) in order to provide the first proxy to relate the ICME orientation with its propagation.}
   {We investigated 22 isolated CME-ICME events in the period 2008-2015. We determined the CME orientation in the near-Sun environment using the following: 1) a 3D reconstruction of the CME with the graduated cylindrical shell (GCS) model applied to coronagraphic images provided by the STEREO and SOHO missions; and 2) an ellipse fitting applied to single spacecraft data from SOHO/LASCO C2 and C3 coronagraphs. In the near-Earth environment, we obtained the orientation of the corresponding ICME using in situ plasma and field data and also investigated the non-radial flow in its sheath region.}
   {The ability of GCS and ellipse fitting to determine the CME orientation is found to be limited to %only distinguishing 
   reliably distinguish only between the high or low inclination of the events. Most of the CME-ICME pairs under investigation were found to be characterized by a low inclination. For the majority of CME-ICME pairs, we obtain consistent estimations of the tilt from remote and in situ data.
   The observed non-radial flows in the sheath region show a greater y direction to z direction flow ratio for high-inclination events, indicating that the CME orientation could have an impact on the CME propagation.}
   {}


    \titlerunning{Determination of CME orientation and consequences for their propagation}
    \authorrunning{Martinic et al.}

   \maketitle
   
%
%-------------------------------------------------------------------

\section{Introduction}
\label{intro}

   Coronal mass ejections (CMEs) are expulsions of magnetic field and plasma from the solar atmosphere into the interplanetary medium. They are known as the main drivers of geomagnetic storms and can cause great damage in the near-Earth environment \citep{Zhang2003}. After a certain distance from the solar surface, the CME dynamics becomes mostly governed by magneto-hydrodynamic "aerodynamic" drag (\citealp{Cargill1996}; \citealp{Vrsnak2001}). This means that CMEs slower than the ambient solar wind are accelerated, while the ones that are faster than the ambient solar wind are decelerated. More recent work on this subject is given by \cite{Temmer2011}, \cite{Vrsnak2013}, and \cite{Sachdeva2015}.

   Coronal mass ejections can be observed remotely using white-light coronagraphs. Coronagraphs situated at different vantage points provide a stereoscopic view and 3D reconstruction of the CME. The graduated cylindrical shell (GCS) model was developed by \cite{Thernisien2006} to perform 3D reconstructions of the CMEs using white-light images from coronagraphs on-board the SOHO and STEREO missions. In the GCS model, the flux-rope (FR) structure is represented with a croissant-like shape that consists of two segments: conical legs and a curved front. Conversely, the cross section of the croissant is circular. Each CME is fully defined by six GCS parameters, these are as follows: 1) the longitude of the apex; 2) the latitude of the apex; 3) the height of the apex; 4) the half-angle, that is a measure of the distance from the leg’s central axis to the apex; 5) the aspect ratio, in other words the measure of the width of the leg; and 6) the tilt, that is the inclination of the FR axis with respect to the solar equator. The GCS implementation for the 3D CME reconstruction as described in \cite{Thernisien2011} has been widely used (e.g., \citealp{Temmer2021}; \citealp{Singh2018}; \citealp{Shi2015}). The orientation of the CME can be obtained using a 2D geometry as well. \cite{Chen1997} suggested that an ellipse can be used to characterize a two-dimensional projection of the CME FR. This was later applied by \cite{Krall2006} and \cite{Byrne2009}, for example, who characterized the observed CME front with an ellipse. By changing the ellipse’s position, axes' length, and tilt, one derives the CME angular width and inclination. To our knowledge, a comparison of the results for CME inclination obtained by these two methods has not been investigated yet.
   
   When crossing the spacecraft, interplanetary coronal mass ejections (ICMEs) show specific signatures. Often a characteristic three-part structure can be observed: shock, sheath, and ejecta/magnetic cloud (MC). The shock arrival is characterized by a sudden increase in the magnetic field, solar wind speed, and temperature. The sheath region is characterized by high turbulence, dense hot plasma of the ambient solar wind, and an interplanetary magnetic field that is compressed and draped around the FR part of the ICME. Also, the sheath region typically shows higher fluctuations in all measured parameters and a smaller radial extension than in the FR part of the ICME \citep{Kilpua2017}. The sheath region in addition shows evidence of non-radial flows (NRFs). \cite{Gosling1987} detected a systematic westward flow in the sheath region and concluded that it is due to the magnetic stress of the Parker spiral acting on the west flank of the ICMEs. Later, \cite{Owens2004} investigated five MCs with relatively uncomplicated upstream NRFs. They found that the deflected flows are more or less parallel to the surface of the MC and that they can be used as a proxy for the local axis orientation and the point of interception of the spacecraft with the ICME. More recently, \cite{Nada2021} performed a statistical research that was focused on NRFs throughout the first 13 years of the STEREO mission. They found that the majority of NRFs are associated with CMEs and that the largest NRFs inside the CME are related to deflections in the sheath region.

   Following the sheath, in situ spacecraft detect the ejected magnetic structure, which occasionally shows clear FR properties. These "magnetic clouds" were first described by \cite{Burlaga1981} and \cite{Klein1982}, and they are characterized by an enhanced and smoothly rotating magnetic field, a depressed proton temperature, and a decreased plasma beta. It has been shown that approximately only one-third of ICMEs show these in situ signatures (\citealp{Gosling1990}; \citealp{Cane2003}). In the first approximation, we can describe the FR part of an ICME as a cylindrical tube that contains a helical magnetic field component which wraps around the tube's central axis and an axial field component which follows the tube's central axis \citep{LUNDQUIST1950}. FRs can have left-handed or right-handed chirality depending on the relative orientation of the helical magnetic field to the axial magnetic filed. Inclination and chirality allows us to classify each FR as one of eight basic types (\citealt{Mulligan1998}; \citealt{Bothmer1998}; \citealt{Palmerio2018}).

  In this study, we analyze the CME orientation obtained by different methods using remote and in situ measurements and the possible impact it could have on the CME propagation. The properties of the interplanetary magnetic field (IMF) and related ambient solar wind differ in the ecliptic and meridional planes \citep{Schwenn2006}. CMEs can have inclinations from extremely low (the ones that lie in the ecliptic plane), to extremely high (the ones that lie in the meridional plane). Consequently, it is reasonable to assume that the interaction of the CME and ambient solar wind are conditioned by the CME's inclination. Non-radial flows in the sheath region are a result of the CME interaction with the ambient solar wind and thus they could indicate different interactions depending on the inclination of the CME propagating through the interplanetary space. The connection between CME inclination and propagation was studied by \cite{Vandas1995} and \cite{Vandas1996}. They performed simulations of magnetic cloud propagation in the inner heliosphere for a high inclination magnetic cloud in the ecliptic plane and for a low inclination magnetic cloud in the meridional plane. They found no significant time arrival difference in these two cases. However, they did not study the propagation of a high inclination magnetic cloud in the meridional plane and a low inclination magnetic cloud in the ecliptic plane.
  
  %                                                One column figure
%----------------------------------------------------------------- 
   \begin{figure}
   \centering
   \includegraphics[width=9.8cm]{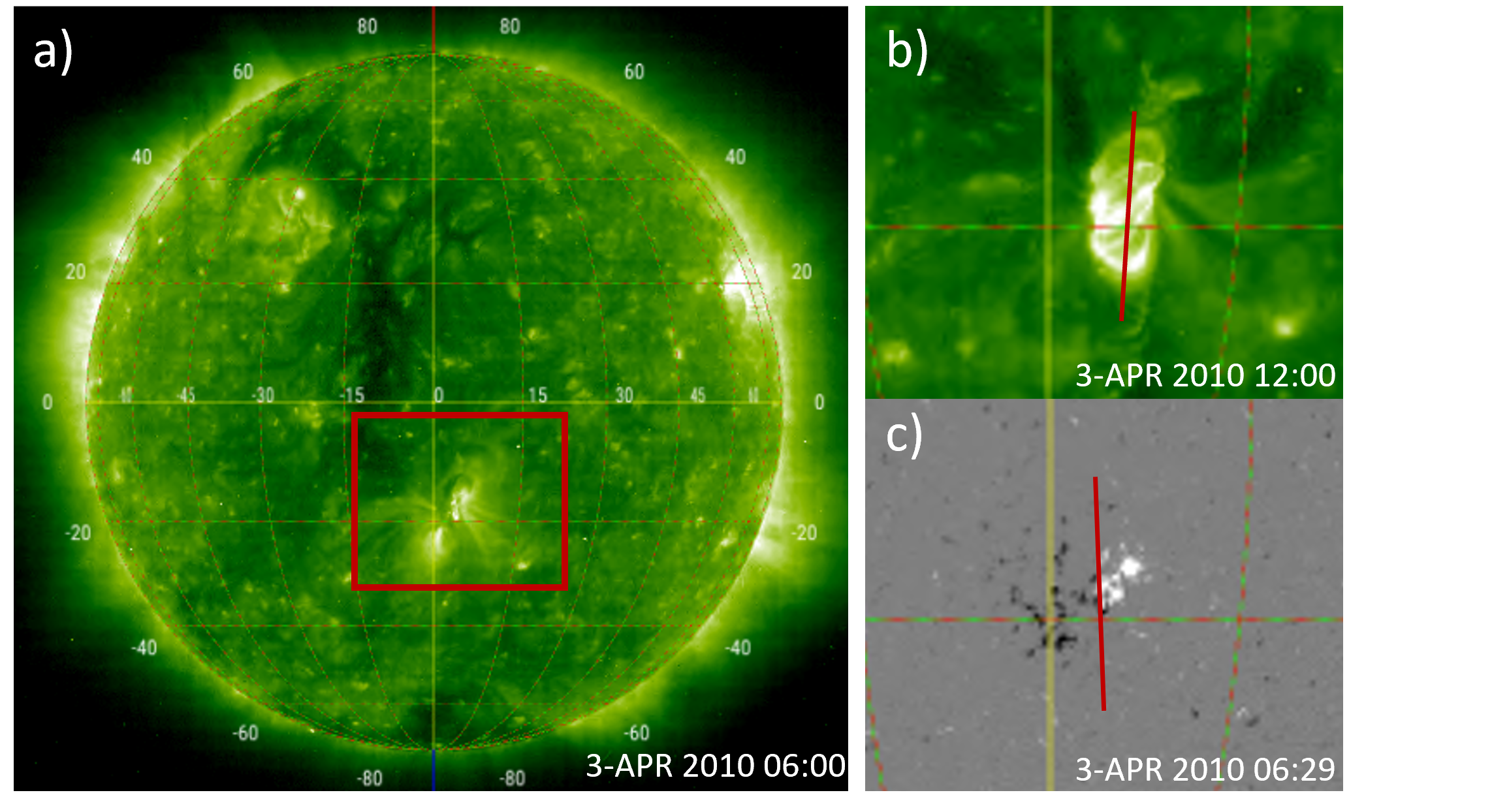}
      \caption{Source region, magnetogram, and low coronal signatures of a CME that occurred on 3 April 2010. a) Source of the eruption (AR 11059) as seen by SOHO/EIT 195 $\AA$ just before the eruption. b) Zoomed-in region indicated by the red rectangle in panel a), showing post flare loops observed by SOHO/EIT 195 $\AA$. c) Same zoomed-in region showing the SOHO/MDI magnetogram. Red lines show tilt estimation.
              }
         \label{fig0}
   \end{figure}
%---------------------------------------------------------------

%--------------------------------------------------------------------
\section{Data and method}
\label{data}

%-------------------------------------- Two column %figure (place early!)
   \begin{figure*}
   \centering
   \includegraphics[width=2\columnwidth]{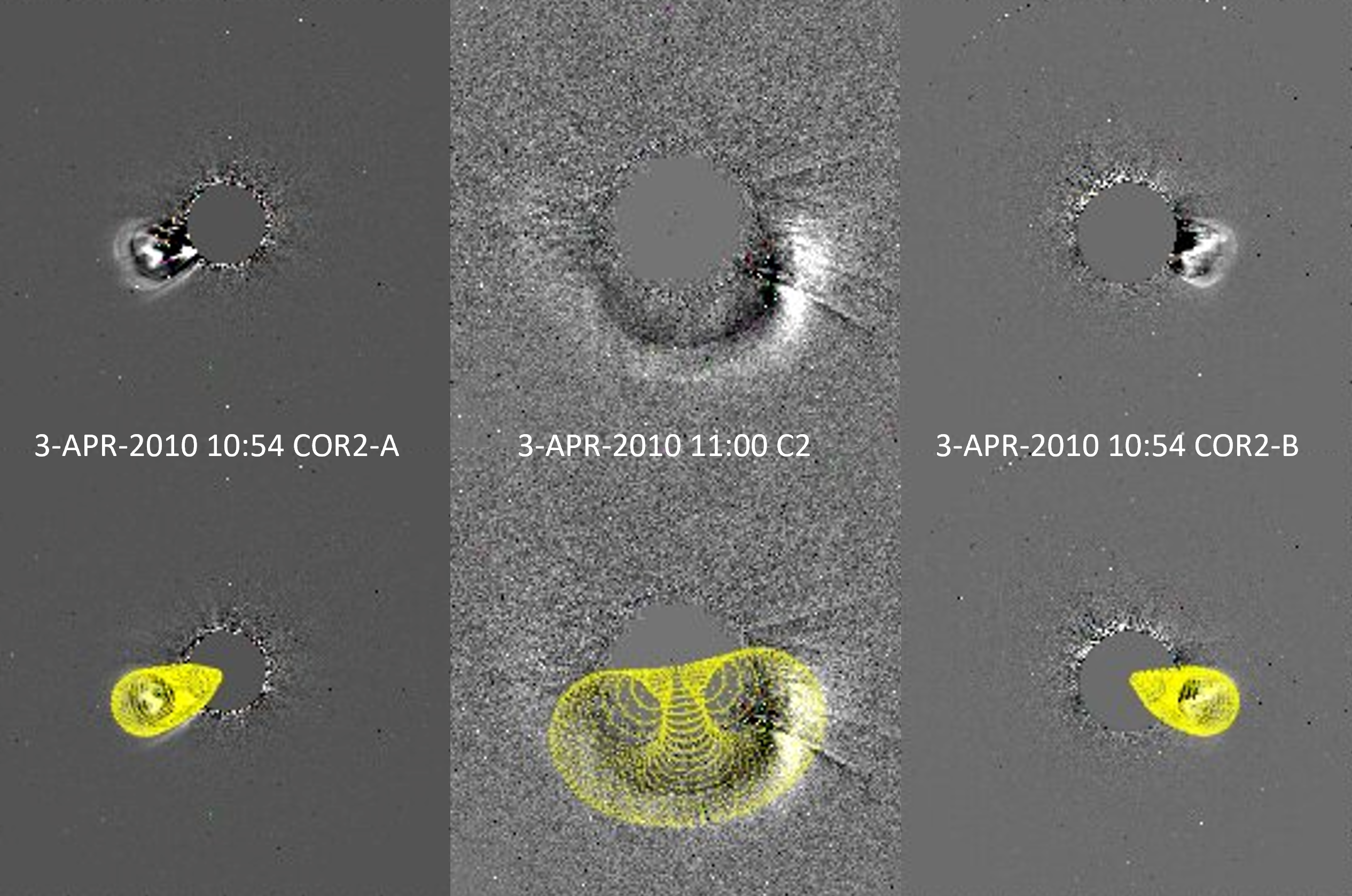}
   \caption{CME that occurred on 3 April 2010. The upper panel shows the running difference images in STEREO(COR2) and SOHO(LASCO-C2), while in the bottom panel the GCS reconstruction is superposed (yellow wire).}%
   \label{fig1}
    \end{figure*}
%---------------------------------------------------

We analyzed CME-ICME orientation using remote and in situ observations. For that purpose, we compiled a list of reliably associated CME-ICME pairs. In order to derive the FR type from the in situ data, we analyzed only the events with a clear MC signature. Moreover, we analyzed only the events that had been remotely observed from at least two vantage points. We performed GCS and ellipse fit analysis only to those events with clearly seen fronts and where image artifacts and/or bright streamers did not affect the front determination. 

We searched for associated CME-ICME pairs in \cite{Palmerio2018} for period 2010-2013, in \cite{Maricic2020} for period 2010-2015, in \cite{Temmer2021} for period 2008-2015, and in \cite{Nitta2017} for period 2010-2016. Altogether, we investigated 63 associated CME-ICME pairs in the time period 2008-2016. Events showing complex (non-MC) in situ signatures that had an unclear CME-ICME association or uncertain GCS reconstruction were discarded. The remaining 22 events have a clear CME-ICME association, clear MC signatures, and the GCS reconstruction was performed with at least two vantage points, that is to say using at least two spacecraft. We note that the majority of the remaining events have a latitude and longitude within $\pm 30^\circ$ from the center of the solar disk as obtained by GCS.  This, along with a clear MC structure observed in situ, indicates a nose hit \citep{Maricic2020}.

The associated CMEs were analyzed in white light observations from the SOHO/LASCO \citep{Brueckner1995} C2 and C3 coronagraphs, STEREO-A(ST-A)/SECCHI, and STEREO-B(ST-B)/SECCHI \cite{Howard2008} COR1 and COR2 coronagraphs. In situ data were provided by the OMNI database \citep{King2005}. The CME orientation was determined using remote observations and two different methods (sections \ref{2.1.1} and \ref{2.1.2}), whereas the ICME orientation was determined using in situ measurements (section \ref{2.2}). Finally, we used in situ measurements to determine the NRFs in the sheath region (section \ref{2.3})
 
 \subsection{Tilt determination in the near-Sun environment}

\subsubsection{Graduated cylindrical shell model}
\label{2.1.1}
We first estimated the CME orientation by performing a 3D FR reconstruction using the GCS model \citep{Thernisien2006}. We used low coronal signatures to better constrain the latitude, longitude, and tilt of the CME. We used JHelioviewer \citep{Muller2017} as a visualization tool to analyze 171, 211, 193, and 304 $\AA$ filtergrams from SDO/AIA \citep{Lemen2012} and SDO/HMI \citep{Scherrer2012} magnetogram data; additionally, when SDO/AIA and/or SDO/HMI data were not available, we used all (E)UV filters from SOHO/EIT \citep{Delaboudini1995} and SOHO/MDI \citep{Scherrer1995} magnetogram data. We searched for post-flare loops (PFLs) whose orientation suggests the orientation of the FR (\citealt{Palmerio2018}; \citealt{Yurchyshyn2008}). Also, we searched for coronal dimmings, sigmoids, and flare ribbons which are known as "by-eye" indicators of the polarity inversion line (PIL) whose orientation roughly matches the orientation of the FR (\citealt{Palmerio2018}, \citealt{Marubashi2015}, \citealt{Mostl2008}). In the case of the quiet Sun eruptions, we searched the position and orientation of the corresponding erupting filament.

An example of how we constrained the latitude, longitude, and tilt of the CME that occurred on 3 April 2010 is shown in figure \ref{fig0}. The pre-eruption SOHO/EIT 195  $\AA$ filtergram, as well as SOHO/MDI, are shown. The active region AR 11059 is marked by the red rectangle on the solar disk as a source of the eruption (Lat $\approx-18^\circ$ and Lon $\approx3^\circ$). In figure \ref{fig0} we also see that the EUV post-flare loops and the PIL in the magnetogram suggest a high-inclination FR as indicated by the red lines determined "by eye".

We performed GCS reconstruction only for the events for which coronagraphic images were available from at least two different vantage points. We reconstructed each event for at least four different heights (i.e., at different times), starting with the lowest heights using coronagraphs COR1-A (STEREO-A), COR1-B (STEREO-B), and C2 (SOHO). We ended the reconstruction at the altitude corresponding to the image of the CME where the front of the FR was last seen unambiguously, using coronagraphs COR2-A (STEREO-A), COR2-B (STEREO-B), and C3 (SOHO).

Figure \ref{fig1} shows the GCS reconstruction (yellow mesh) for event 2 from the list (table \ref{tab1}, the CME occurred on 3 April 2010). This is an example of a low inclination event and the GCS reconstruction was obtained using coronagraphic images from three different vantage points. 

We see that the inclination derived from GCS greatly differs from the inclination estimated from post-flare loops and magnetogram for the same event. This is not unusual since the evidence for rotation and deflection in the low and middle corona has been presented many times (\citealt{Fan2004}; \citealt{Green2007}; \citealt{Lynch2009};  \citealt{Vourlidas2011}; \citealt{Kay2017}). However, it is beyond the scope of our study to further analyze possible rotations of each event. We emphasize once more that the priority was given to the GCS-obtained inclination and that orientation estimation with low coronal signatures and magnetogram data was taken only as a possible constraint.

\subsubsection{Ellipse fit}
\label{2.1.2}

\begin{figure}
   \centering
   \includegraphics[width=9cm]{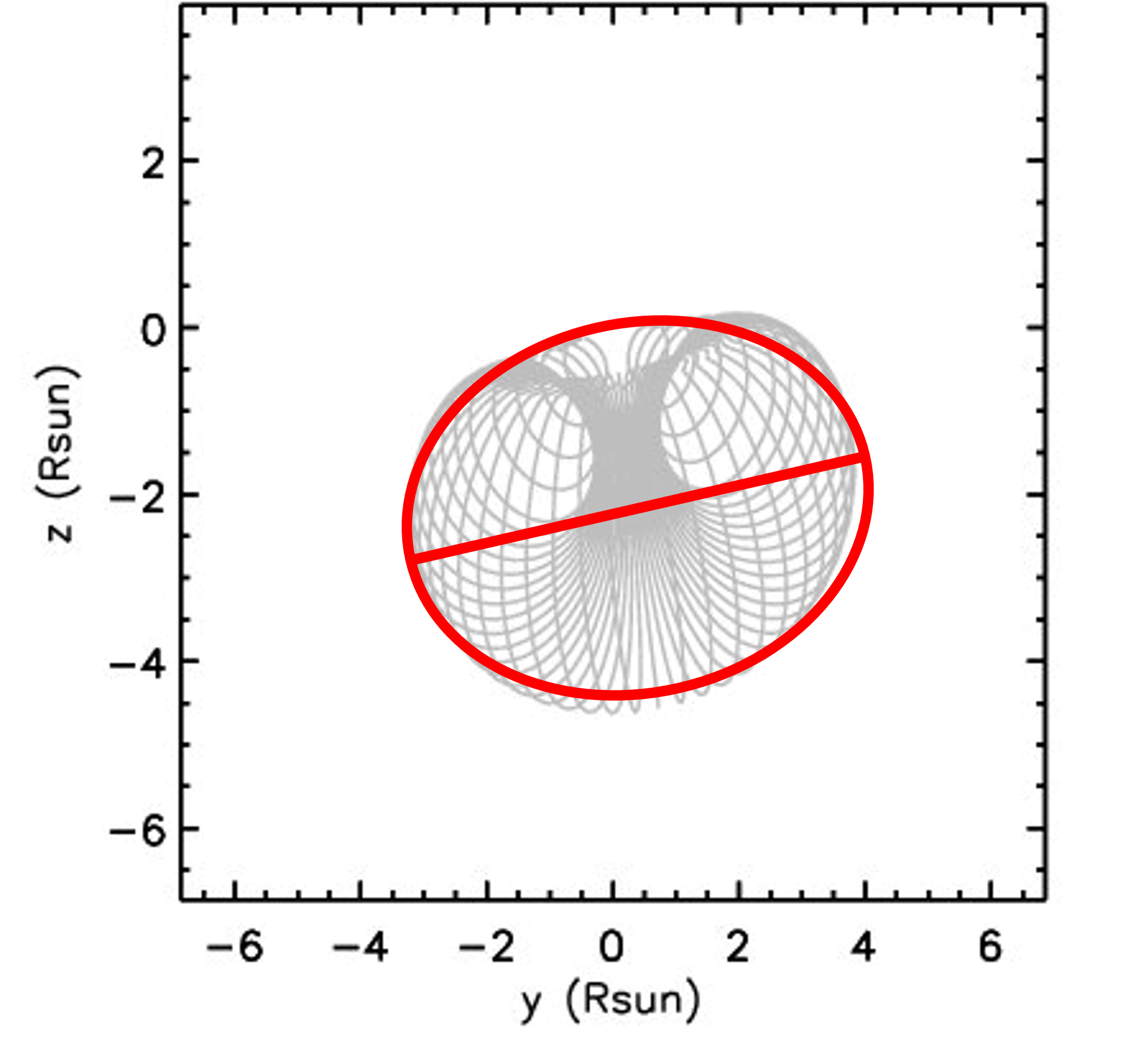}
      \caption{Projection of the GCS croissant of the 3 April 2010 CME in the yz plane of the Heliocentric Earth EQuatorial (HEEQ) coordinate system (i.e., Earth view). A possible ellipse representation is marked by the red line along with the ellipse's major axis. 
              }
         \label{fig2}
   \end{figure}

 The projection of an Earth-directed, GCS-obtained croissant, in the yz plane (Earth view) of the Heliocentric Earth EQuatorial (HEEQ) coordinate system can be approximated with an ellipse. An example of the ellipse approximation of the GCS-obtained croissant of the event that occurred on 3 April 2010 is shown with the red-colored ellipse in figure \ref{fig2}. Led by this idea, we performed the ellipse fit on data provided by C2 and C3 coronagraphs on board the SOHO spacecraft. The front of each Earth-directed CME observed with C2 and C3 coronagraphs are represented with an ellipse and the inclination of the ellipse's major axis to the equator is taken as the CME tilt. 
%                                                One column figure
%----------------------------------------------------------------- 
%---------------------------------------------------------------

%                                           One column figure
%----------------------------------------------------------------- 
   \begin{figure}
   \centering
   \includegraphics[width=9cm]{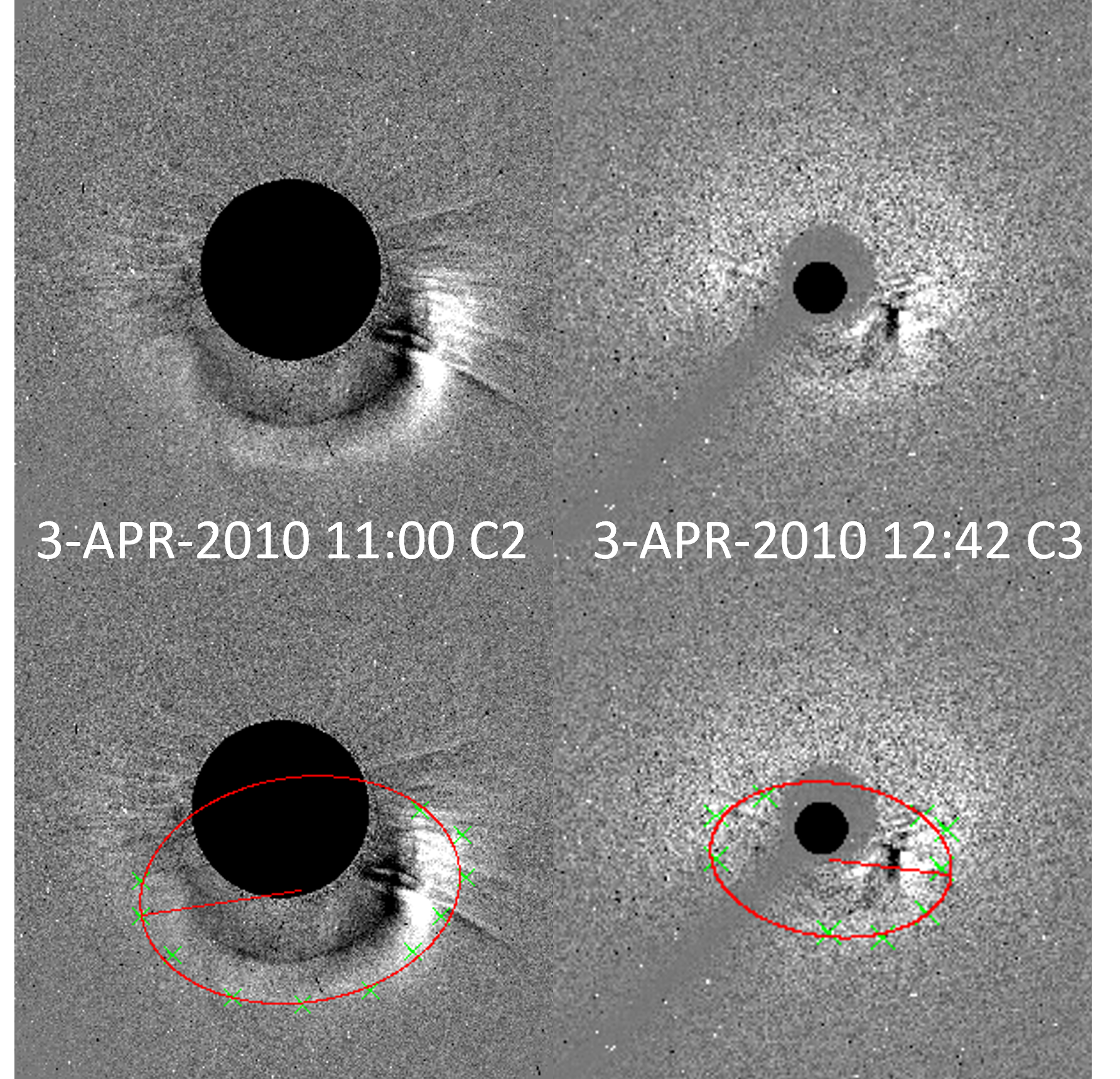}
      \caption{CME that occurred on 3 April 2010. The upper panel shows the running difference images in SOHO (LASCO-C2 and LASCO-C3). The bottom left panel shows the results of the ellipse C2  fitting at the same time when the GCS was performed. The bottom right panel shows the results of the ellipse fitting when using data from C3. The ellipse is represented with the red line and green crosses mark the points outlined on the CME front used to obtain the fit.}
         \label{fig3}
   \end{figure}
%---------------------------------------------------------------

Figure \ref{fig3} shows the ellipse fit results for the same event that is shown in figures \ref{fig1} and \ref{fig2}. The panel on the left (figure \ref{fig3}) shows the ellipse fit obtained with LASCO(C2) at the same time as the GCS reconstruction. The panel on the right (figure \ref{fig3}) shows the ellipse fit obtained with LASCO(C3) for the same event, but at a later time. It is important to note that the relative size of the occulting disk compared to the overall size of the observed structure (CME) may influence the ellipse fitting. The greater the size of the occulting disk of the coronagraph compared to the size of the CME, the harder it is to perform the fit. Understandably, this is more pronounced when doing the ellipse fit with the LASCO(C2) than with the LASCO(C3) images. On the other hand, the CME front for some events becomes faint in the C3 field of view (FOV) and thus is more difficult and unreliable to fit. Therefore, we performed an ellipse fit using both C2 and C3 data. The robustness of two different methods applied to different data sets used to determine the tilt in the near-Sun environment (GCS, ellipse-C2, and ellipse-C3) is presented and discussed in section \ref{sec3}.

It is worth emphasizing that we did not introduce the ellipse fit method in order to increase the reliability of the GCS reconstruction, but rather to compare the results of the two methods. GCS and the ellipse fit both use morphological features of a CME for the reconstruction, but GCS uses a 3D geometry (croissant) whereas the ellipse fit uses a 2D geometry (ellipse). Thus, we can not a priori know whether the methods will give similar tilt results. The main motivation for testing this is to provide a reference for future work so that we can study a larger statistical sample of CME-ICME associations by searching through the whole SOHO era.

\subsection{Tilt determination in the near-Earth environment}
\label{2.2}

As a next step, we determined the tilt of the ICME in the near-Earth environment using in situ data obtained from the WIND and ACE spacecraft and provided through the OMNI database \citep{King2005}. First, we determined the ICME and MC boundaries. In order to achieve consistency as the main criterion for ICME arrival, we used the sudden increase in the magnetic field, density, temperature, and velocity to mark the ICME shock or sheath arrival. To consistently determine MC boundaries, we used magnetic field smooth rotation as the main criteria for all studied events. The determined end of the MC was taken as the end of the ICME as well.
 
We determined if the event is dominantly high or low inclined from its characteristic in situ signatures. FRs that have their central axis more or less parallel to the ecliptic plane are called low-inclination FRs (the $B_z$ component represents the helical field and thus its sign changes as the FR is crossed). FRs that have their central axis more or less perpendicular to the ecliptic plane are called high-inclination FRs \citep[the $B_z$  component represents the axial field and thus its sign does not change as the FR is crossed, see e.g.,][]{Palmerio2018}. Early work by \cite{Mulligan1998} and \cite{Bothmer1998} suggested the existence of eight different types of the magnetic FR with different magnetic configurations of MCs which can be observed during the cloud's passage. The FR-type determination in the near-Earth environment according the abovementioned "eight-type" classification also allows us to distinguish between a high (ESW, ENW, WSE, and WNE) and low (NES, NWS, SEN, and SWN) inclination. The determination of 22 observed events according to this eight-type classification is shown in the last column of table \ref{tab2}.  

We note that this classification of a FR only provides us with information on whether it has high or low inclination. There are various FR
reconstruction methods one could apply to in situ data to obtain the value of the FR tilt. However, it was shown by \cite{Al-Haddad2013} that the determination of the value of the FR tilt can be quite unreliable. They performed four different reconstruction and fitting methods on 59 ICMEs observed in situ. All four methods gave an orientation of the FR axis within $\pm 45^\circ$ for only one event. They also found that other results, besides inclination, obtained with different techniques usually did not match. Therefore, we constrained our estimation to high and low inclination from the in situ data for each event.

%                                                One column figure
%----------------------------------------------------------------- 
   \begin{figure}
   \centering
   \includegraphics[width=9cm]{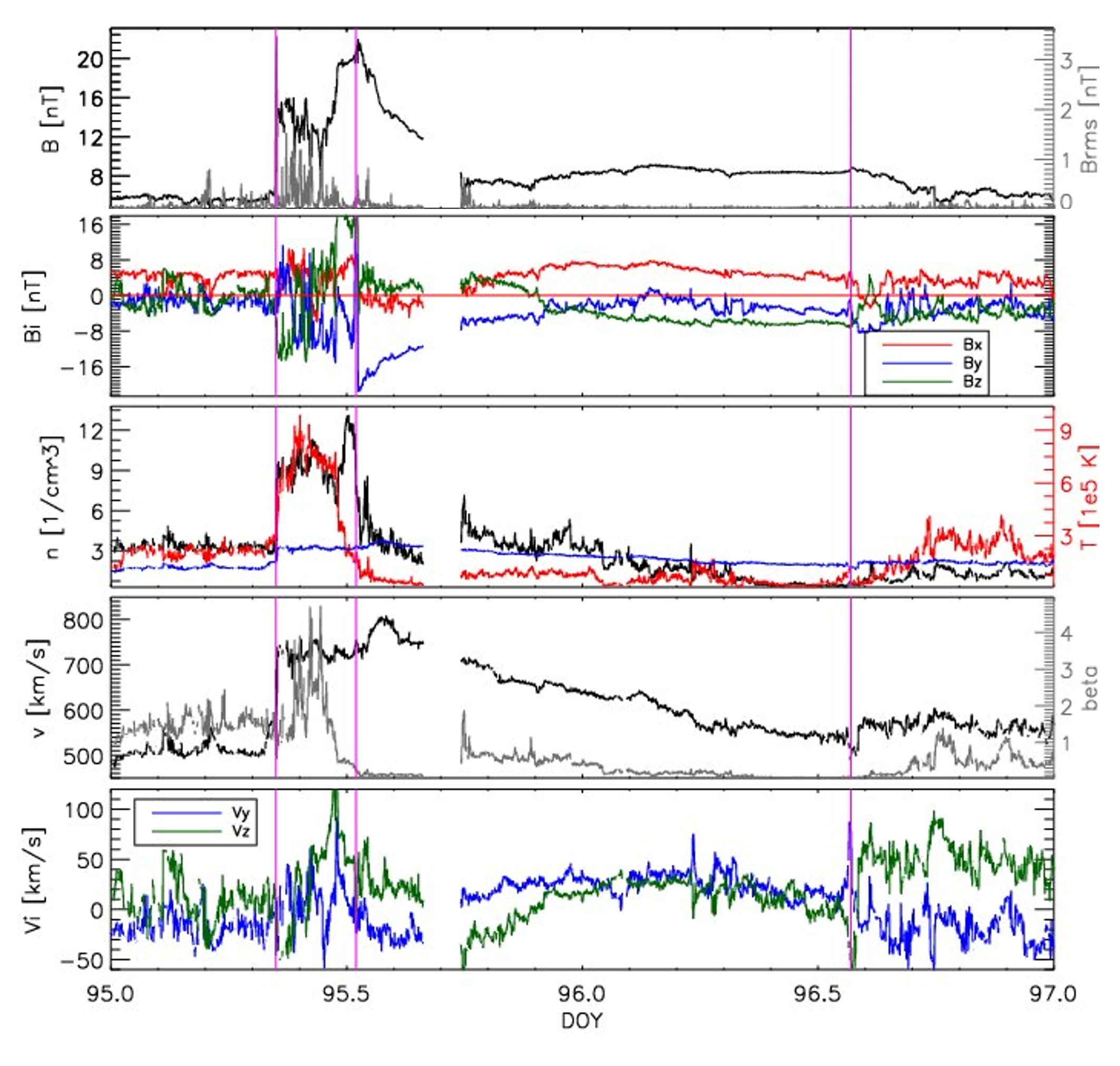}
      \caption{ICME observed in situ measurements on 5 April 2010. The parameters shown, from top to bottom, are the magnetic field magnitude (black) and magnetic field fluctuations (gray); magnetic field components in the GSE coordinate system (red, $B_x$; blue, $B_y$; green, $B_z$ ); the proton density (black) and temperature (red) along with the calculated expected temperature (blue); the solar wind speed (black) and plasma beta (gray); and finally the y (blue) and z (green) velocity component in the GSE coordinate system.  The vertical magenta lines indicate, from left to right, shock-sheath arrival, the leading edge of the MC, and finally the trailing edge of the MC which is the same as the end of the whole ICME.}
      \label{fig4}
   \end{figure}
%---------------------------------------------------------------

Figure \ref{fig4} shows the same CME that was launched from the Sun on 3 April 2010, as seen two days later in the in situ data. This event was classified as a low inclined event due to clear rotation of the $B_z$ component in the marked MC region.

\subsection{Non-radial flows in the sheath region} 
\label{2.3}

We analyzed NRFs in the sheath region of the in situ observed ICMEs. One would expect NRFs to be locally, approximately parallel to the surface of the ICME, thus the properties of the NRFs should reflect the ICME geometry \citep{Owens2004}. Since we considered only CME-ICME pairs that were approximate nose hits, we might expect low inclined ICMEs to have more pronounced NRFs in the $\pm$ z direction of the GSE coordinate system in comparison to high inclined ICMEs. We derived the NRFs in the z direction and y direction of the GSE coordinate system by calculating the average values of $v_z$ and $v_y$ absolute values in the sheath region, respectively. In order to test the hypothesis that low inclined events have more profound flows in the z direction, we define the NRF ratio $\theta$ as follows:

\begin{align*}
    \theta&=\frac{\overline{|v_y|}}{\overline{|v_z|}},
\end{align*}

\noindent where $\overline{|v_y|}$ and $\overline{|v_z|}$ are the mean values of the magnitude of the velocity in the sheath region in the y direction and z direction, respectively. 

\section{Results and discussion}
\label{sec3}

Table \ref{tab1} lists 22 events and the results of the GCS reconstruction: longitude, latitude, tilt, aspect ratio, and half-angle. The first column shows the ordinal number of the event and the upper index indicates from which CME-ICME list the association was taken; P stands for \cite{Palmerio2018}, M stands for \cite{Maricic2020}, T stands for \cite{Temmer2021}, and N stands for \cite{Nitta2017}. The second column shows the time when the GCS reconstruction was performed.

%-------------------------------------------------------------
%                                             Two column Table 
%-------------------------------------------------------------
%
\begin{table*}
\caption{Results of the GCS modeling. We provide the event number with an indication from where the CME-ICME association was taken, reconstruction time, Stonyhurst longitude, latitude, CME tilt,  aspect ratio, and half-angle.}             
\label{tab1}      
\centering     
\newcolumntype{d}[1]{D{.}{\cdot}{#1} }
\begin{tabular}{l c r r r r r r }     % 7 columns 
\hline\hline       
                      % To combine 4 columns into a single one 
NO  & Reconstruction time & Long[$^\circ$] & Lat[$^\circ$] & Tilt[$^\circ$] & Height [$R_s$] & Aspect ratio & Half-angle[$^\circ$] \\ 
\hline                    
 $1^{T}$   & 2008-12-12 11:54 & 0 & 6 & 38 & 11.8 & 0.28 & 18 \\ 

 $2^{M,T}$   & 2010-04-03 10:55 & 5 & -26 & 10 & 7.7 & 0.35 & 30 \\

 $3^{P}$  & 2010-05-23 21:54 & 0 & 5 & 35 &12.1 & 0.27 & 21\\
 
 $4^{T,N}$   & 2010-06-16 19:24 & 351 & 5 & -23 & 10.8 & 0.29 & 17\\
 
 $5^{T,N}$   & 2011-01-30 17:39 & 0 & -17 & 0 & 10.9 & 0.35 & 30\\ 

 $6^{P,M,N}$   & 2011-03-25 10:39 & 333 & -3 & -23 & 6.6 & 0.21 & 17 \\

 $7^{N}$   & 2011-05-25 13:54 & 3 & 9 & 58 & 9.2 & 0.13 & 13 \\
 
 $8^{P,T}$   & 2011-06-02 08:39 & 352 & -5 & 17 & 7.3& 0.35 & 30 \\
 
 $9^{P,M,T}$   & 2011-09-14 01:24 & 17 & 22 & -36 & 7.8 & 0.34 & 25\\ 

 $10^{P,T}$   & 2012-01-19 15:10 & 323 & 48 & 90 & 3.9 & 0.32 & 29 \\

 $11^{P,M}$   & 2012-05-12 00:48 & 330 & -8 & 90 & 10.1 & 0.27 & 18\\
 
 $12^{P,M,T}$   & 2012-06-14 14:08 & 0 & -20 & 38  & 4.2 & 0.31 & 18 \\
 
 $13^{P,T,N}$   & 2012-10-05 07:24 & 12 & -14 &46 & 14.9 & 0.24 & 31\\
 
 $14^{P}$   & 2012-10-09 07:39 & 1 & 5 & 4 & 13.7 & 0.32 & 29 \\ 

 $15^{T}$   & 2012-11-09 16:24 & 356 & -13 & 19 & 9.1 & 0.33 & 30 \\

 $16^{P}$   & 2013-01-13 15:54 & 1 & -1 & -6 & 12.4 & 0.34 & 11 \\
 
 $17^{P,M,T}$  & 2013-04-11 08:24 & 343 & -7 & 66 & 9.0 & 0.30 & 21 \\
 
 $18^{N}$  & 2013-06-02 22:54 & 0 & -3 & 12 & 12.1 & 0.35 & 28 \\
 
 $19^{P,M,T}$   & 2013-07-09 15:12 & 1 & 2 & 0 & 12.3 & 0.46 & 31\\ 

 $20^{M,T}$   & 2013-09-29 22:39 & 7 & 27 & -67 & 6.6 & 0.32 & 34  \\

 $21^{P,T}$   & 2014-08-15 17:48 & 9 & 15 & -52 & 12.7 & 0.22 & 20 \\
 
 $22^{N}$   & 2016-10-09 04:54 & 0 & 10 & -23& 8.9 & 0.35 & 31 \\
\hline                  
\end{tabular}
\end{table*}
%
%-------------------------------------------------------------

Due to the subjectiveness of the GCS reconstruction, we compared these results to the tilt results obtained using GCS by \cite{Temmer2021}, \cite{Sachdeva2019}, and the \href{https://www.helcats-fp7.eu/catalogues/wp3_kincat.html}{HELCATS catalog}, shown in table \ref{tab2}. Also, in table \ref{tab2} the tilt results obtained by the ellipse fitting performed on coronagraphic images provided by the SOHO/LASCO C2 and C3 mission are shown. Events 5, 14, 16, and 18 have very faint leading-edge fronts in the C2 and C3 FOV, so the GCS reconstruction was obtained only using STEREO-A and STEREO-B data, and the ellipse fit was not performed at all. Events 4, 6, 7, 19, and 22 lost the clear leading edge front from the C2 to C3 FOV, so we were not able to perform the ellipse fitting on C3 coronagraphic images.

%-------------------------------------------------------------
%                                             Two column Table 
%-------------------------------------------------------------
%
\begin{table*}
\caption{Comparison of CME tilt results from different techniques and studies. The first column shows the event number with indications wherefrom the CME-ICME association was taken. Next are the results for the tilt obtained by GCS (same as in table \ref{tab1}), ellipse fit C2 tilt results, ellipse fit C3 tilt results, tilt results from \cite{Temmer2021}, tilt results from \cite{Sachdeva2019}, and tilt results from the \href{https://www.helcats-fp7.eu/catalogues/wp3_kincat.html}{HELCATS catalog}. We note that $L$ stands for a low inclination result, $H$ stands for a high inclination result, and $L/H$ stand for a tilt result that could be considered low and high.}             
\label{tab2}      
\centering       
\newcolumntype{d}[1]{D{.}{\cdot}{#1} }
\begin{tabular}{l r r r r r r}     % 7 columns 
\hline\hline       
  NO  & GCS & Ellipse C2 & Ellipse C3 & Temmer+2021 & Sachdeva PHD & HELCATS   \\
 \hline
 $1^{T}$   & 38$^\circ$(L) & 19$^\circ$(L) & -23$^\circ$(L) & 51$^\circ$(H) & / & 51$^\circ$(H)  \\ 

 $2^{M,T}$   & 10$^\circ$(L) & 20$^\circ$(L) & -7$^\circ$(L) & 2$^\circ$(L) & 22$^\circ$(L) & /  \\

 $3^{P}$  & 35$^\circ$(L) & 27$^\circ$(L) & 29$^\circ$(L) & / & / & -10$^\circ$(L)  \\
 
 $4^{T,N}$   & -23$^\circ$(L) & 19$^\circ$(L) & / & -55$^\circ$(H) & -15$^\circ$(L) & -8$^\circ$(L)  \\
 
 $5^{T,N}$   & 0$^\circ$(L) & / & / & -20$^\circ$(L) & / & -20$^\circ$(L)  \\ 

 $6^{P,M,N}$   & -23$^\circ$(L) & 19$^\circ$(L) & / & / & 9$^\circ$(L) & 0$^\circ$(L)  \\

 $7^{N}$   & 58$^\circ$(H) & 50$^\circ$(L/H) & / & / & / & /  \\
 
 $8^{P,T}$   & 17$^\circ$(L) & -49$^\circ$(L/H) & -19$^\circ$(L) & -55$^\circ$(H) & /& 55$^\circ$(H)  \\
 
 $9^{P,M,T}$   & -36$^\circ$(L) & -32$^\circ$(L) & -39$^\circ$(L) & -6$^\circ$(L) & / & -6$^\circ$(L)  \\ 

 $10^{P,T}$   & 90$^\circ$(H) & -22$^\circ$(L) & -43$^\circ$(L/H) & 90$^\circ$(H) & 90$^\circ$(H) & /  \\

 $11^{P,M}$   & 90$^\circ$(H) &  -14$^\circ$(L) & -21$^\circ$(L) & / & / & / \\
 
 $12^{P,M,T}$   & 38$^\circ$(L) & 26$^\circ$(L) & 30$^\circ$(L) & 67$^\circ$(H) & -87$^\circ$(H) & /  \\
 
 $13^{P,T,N}$   & 46$^\circ$(L/H) & 45$^\circ$(L/H) & 72$^\circ$(H) & 41$^\circ$(L/H) & 37$^\circ$(L) & 54$^\circ$(H)  \\
 
 $14^{P}$   & 4$^\circ$(L) & / & / & / & / & / \\ 

 $15^{T}$   & 19$^\circ$(L) & 31$^\circ$(L) & 22$^\circ$(L) & 6$^\circ$(L) & 7$^\circ$(L) & 22$^\circ$(L)  \\

 $16^{P}$   & -6$^\circ$(L) & / & / & / & / & /  \\
 
 $17^{P,M,T}$  & 66$^\circ$(H) & -37$^\circ$(L) & 41$^\circ$(L/H) & 66$^\circ$(H) & 90$^\circ$(H) & /  \\
 
 $18^{N}$  & 12$^\circ$(L) & / & / & / & / & /  \\
 
 $19^{P,M,T}$   & 0$^\circ$(L) & 15$^\circ$(L) & / & /& / & /  \\ 

 $20^{M,T}$   & -67$^\circ$(H) & -54$^\circ$(H) & -66$^\circ$(H) & 90$^\circ$(H) & 90$^\circ$(H) & -67$^\circ$(H)   \\

 $21^{P,T}$   & -52$^\circ$(H) & -40$^\circ$(L/H) & -44$^\circ$(L/H) & / & / & /  \\
 
 $22^{N}$   & -23$^\circ$(L) & -25$^\circ$(L) & / & -23$^\circ$(L) & / & /  \\
\hline                  
\end{tabular}
\end{table*}
%
%-------------------------------------------------------------

From table \ref{tab1} we can see that there are $50\%$ more low than high inclined events in the near-Sun environment. From table \ref{tab2} it is obvious that different methods give different results for the same event and we cannot say that a certain event has a certain tilt. In order to at least conclude as to whether an event is dominantly low or high inclined, we added an $L$($H$) mark near each tilt result presented in table \ref{tab2}.We note that  $L$ was given for results $\tau \in [\pm0^\circ,\pm44^\circ]$ and $H$ was given for results $\tau \in [\pm45^\circ,\pm90^\circ]$. The events with $\tau \in [\pm40^\circ,\pm50^\circ]$ are indicated as $L/H$ because those could be considered as both high and low inclined.

 From now on, the results obtained with different methods that are the same in terms of high and low inclination are referred to as "consistent" results. We also emphasize that events marked as $H/L$ for a certain method were not taken into account for the statistics as either high or low inclination events. When comparing our GCS tilt result with C2- and C3-ellipse fitting, we see that for seven out of 13 events ($54\%$) for which we were able to perform GCS, ellipse-C2, and ellipse-C3 fitting methods, the obtained tilt was consistent. Moreover, the GCS and C2-ellipse fit gave consistent results for 12 out of 18 events ( $67\%$). When comparing the GCS and C3-ellipse fit, we obtained consistent results for eight out of 13 events ($62\%$). We note that C2- and C3-ellipse fitting gave consistent results for nine out of 13 events ($70\%$). In regards to a comparison between our results, GCS results, those of \cite{Temmer2021} as well as \cite{Sachdeva2019}, and the \href{https://www.helcats-fp7.eu/catalogues/wp3_kincat.html}{HELCATS catalog}, we see that in 11 out of 15 ($73\%$) events our GCS results were consistent with the majority of the studies listed above. In four out of 15 ($27\%$) events, our GCS results differed from the majority of other research results listed in table \ref{tab2}, again in the scope of high and low inclination.

To summarize, we see that in the majority of events, the C2- and C3-ellipse fit provide the same results. Also, the majority of these C2- and C3-ellipse fit results are in agreement with the GCS results. 
However, the methods are robust only in the scope of determining whether a certain event has either a high or low inclination, but not in terms of a specific value.

When determining the FR type using the magnetic field components' rotation from in situ data, we found that the majority of events have a low inclination (see table \ref{tab3}). Only eight out of 22 events ($36\%$) were considered as high inclined.

When comparing the results for inclination derived with GCS (table \ref{tab2}) and the results for inclination from in situ data (table \ref{tab3}), we found that 14 out of 22 ($63\%$) events have a consistent tilt estimation from remote and in situ measurements. We also see that there are slightly more events that were classified as low inclined from remote observations and as high inclined from in situ data ($18\%$) than vice versa ($14\%$). This is in agreement with the results from \cite{Xie2021}. They compared the orientations of 102 CMEs at a near-Sun environment obtained from the EFR model \citep{Krall2006} with the orientations obtained at L1 using a simple cylindrical force-free FR model. They found that only $25\%$ of the studied events show rotations greater than $40^\circ$ and that the majority of these rotational events occurred within the COR2 FOV, that is to say the middle corona. 

 As stated above, nearly one-third of the events under study have an inconsistent inclination as derived remotely (GCS) and in situ. We identify four possible reasons for this: 1) wrong association; 2) wrong tilt estimation remote; 3) wrong tilt estimation in situ; and 4) real tilt angle change during propagation (outside of the COR2 FOV).

\noindent During the event selection, we took care to consider only events with good CME-ICME associations, so this is unlikely to be the cause of the inconsistency. The intrinsic features of methods for tilt determination come into question as well, especially the difference between remote sensing and in situ tilt determination. Namely, we are looking at the global structure of the CME remotely, while in situ we can see only local features of the FR across the spacecraft crossing line. Furthermore, it has been shown that CME rotations occur frequently during the eruption and in the first few solar radii of the CME propagation (\citealt{Fan2004}; \citealt{Green2007}; \citealt{Lynch2009};  \citealt{Vourlidas2011}; \citealt{Kay2017}), but some authors have also presented evidence of CME rotation outside of the corona \citep{Isavnin2014}. From this perspective, we can conclude that rotations in interplanetary space are possible, but not very likely. However, it is beyond the scope of this study to explain the tilt inconsistency as seen remotely and in situ.

\begin{figure}
   \centering
   \includegraphics[width=9cm]{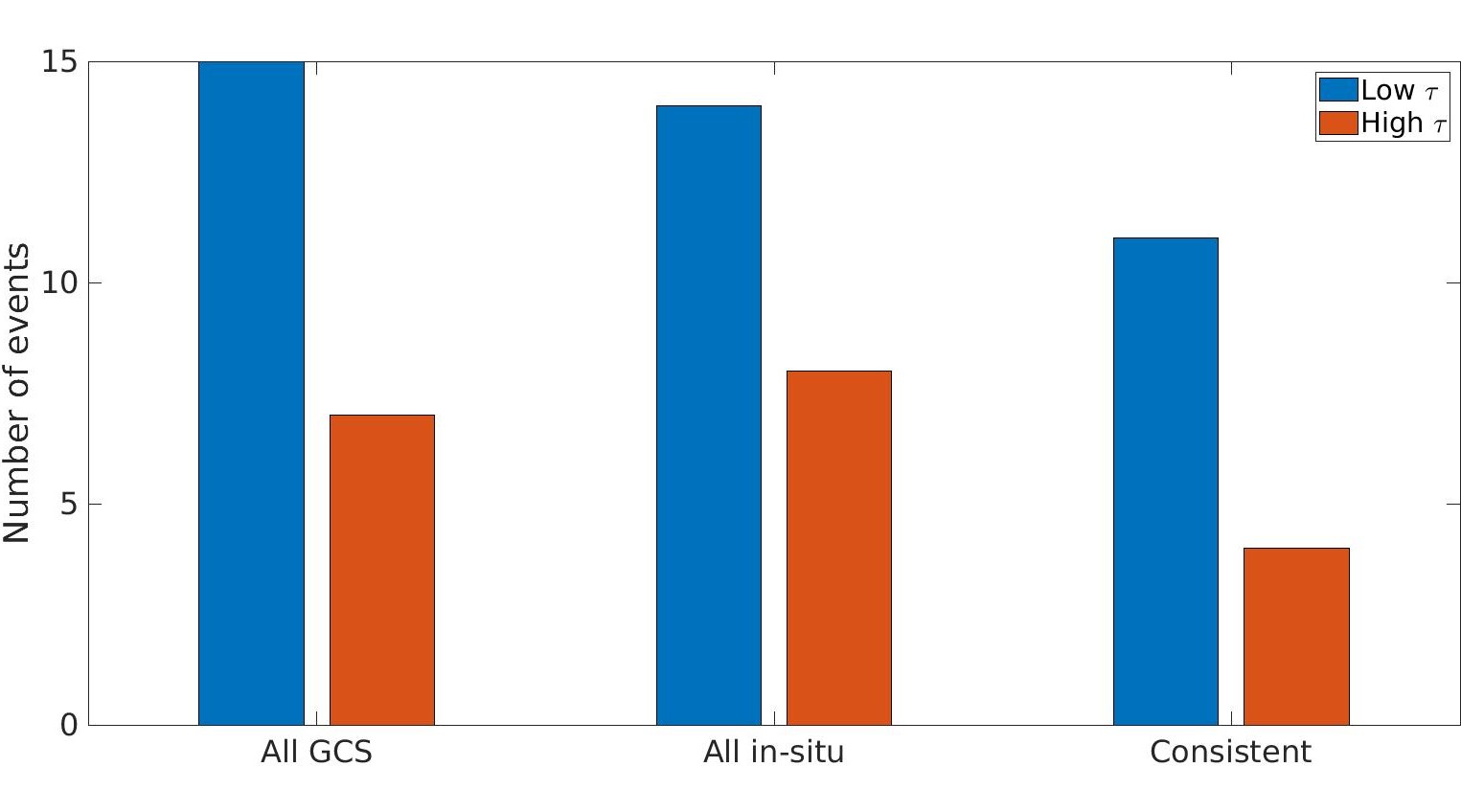}
      \caption{Depiction of how many high and low inclination events were observed using the GCS model in the near-Sun environment, how many high and low inclination events were observed using in situ data in the near-Earth environment, and how many consistent events were classified as low and high inclination.}
         \label{fig5}
   \end{figure}

The bar chart in figure \ref{fig5} shows how many high and low inclination events were observed using the GCS model in the near-Sun environment, how many high and low incline events were observed using in situ data in the near-Earth environment, and how many events have consistently measured low and high tilts in both remote and in situ measurements. We found that in all three cases, the majority of events are characterized by a low inclination.

%                                                One column figure
%----------------------------------------------------------------- 
%---------------------------------------------------------------

%                                                One column figure
%----------------------------------------------------------------- 
   \begin{figure}
   \centering
   \includegraphics[width=9cm]{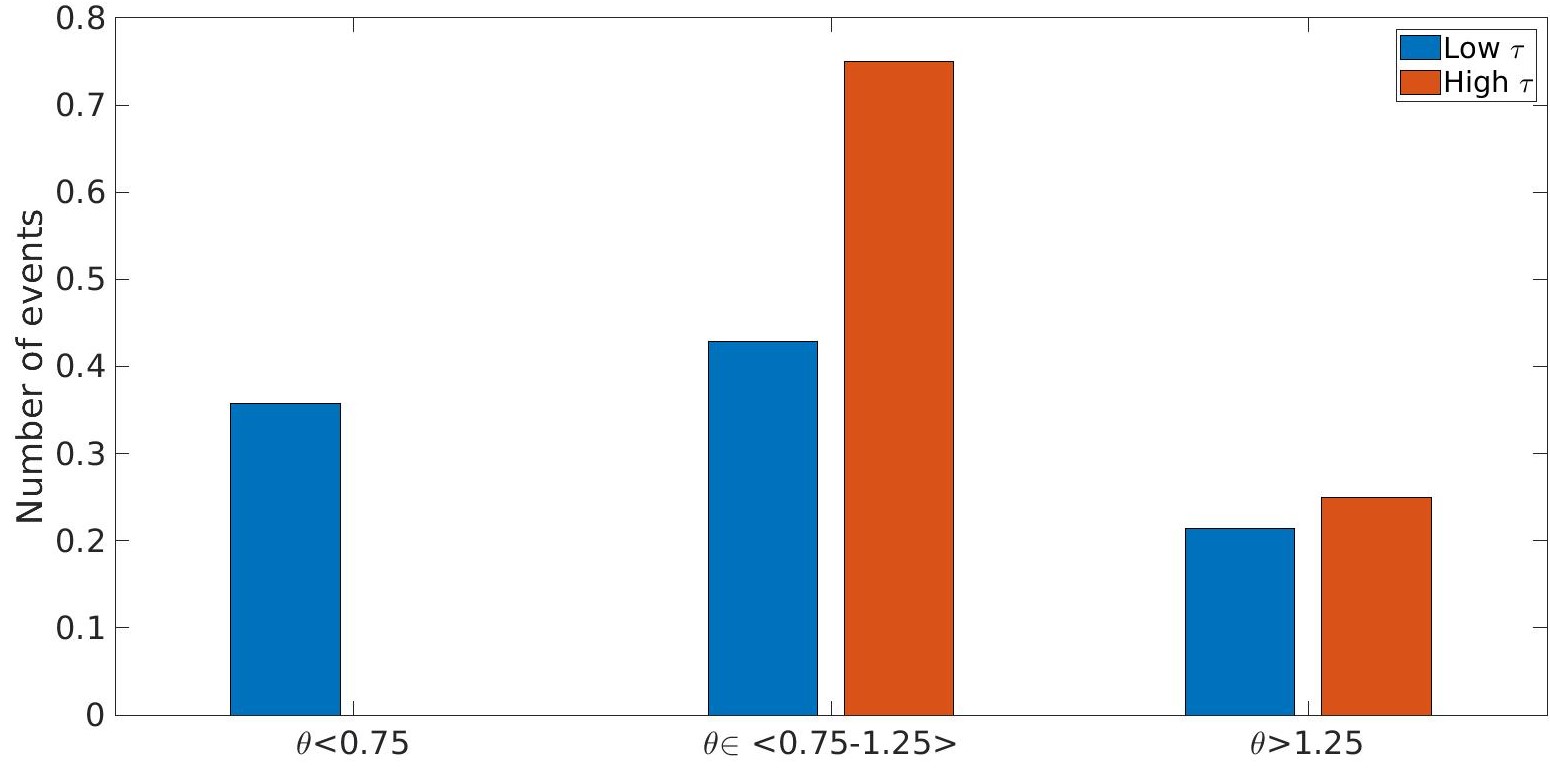}
   \label{Fig6}
      \caption{Occurrence frequency of events with respect to $\theta$ ratio. The events are divided into three groups, $\theta<0.75$, $\theta\in <0.75,1.25>$, and $\theta>1.25$. The occurrence frequency for high inclination events is shown in orange, and for low inclination it is shown with blue bars.}
         \label{fig6}
   \end{figure}
%---------------------------------------------------------------

%-------------------------------------- Two column %figure (place early!)
   \begin{figure*}
   \centering
   \includegraphics[width=2\columnwidth]{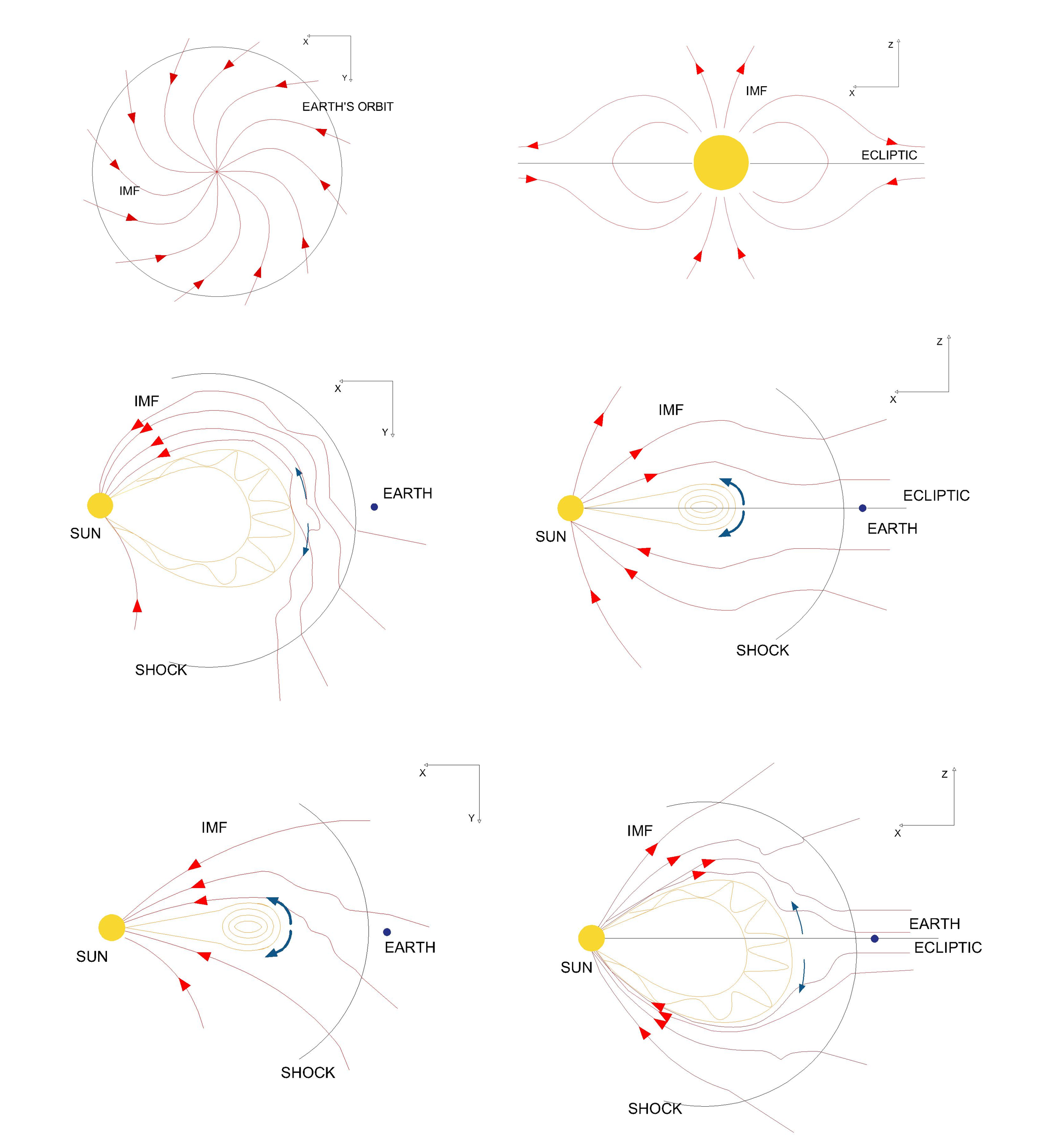}
   \caption{Idealized IMF in the ecliptic and meridional plane (top panels) and its interaction with embedded CME with low (middle panels) and high (bottom panels)  inclination shown schematically. Each panel has a 2D coordinate system drawn in the upper right corner that notes either a $yx$ plane or an $xz$ plane of the GSE coordinate system. The panels also marks the NRF with blue arrows; its width and length suggest the pronouncement of the flow. The figure was adapted from \cite{Gosling1987}.}%
   \label{fig7}
    \end{figure*}
    
%---------------------------------------------------

\begin{table*}
\caption{Results of the in situ event determination are listed. From left to right, the table shows the following: leading-edge (LE) date appearance, LE day of the year (DOY) time appearance, MC DOY time appearance, trailing-edge (TE) DOY time appearance, flux-rope (FR) type, and the classification according to the tilt of the event.}             
\label{tab3}      
\centering  
\newcolumntype{d}[1]{D{.}{\cdot}{#1} }
\begin{tabular}{l c r r r c c r}     % 8 columns 
\hline\hline       
 NO  & LE date & LE DOY & MC DOY & TE DOY & FR Type & Inclination & $\theta$   \\
 \hline
 $1^{T}$   & 2008-12-16 & 351.48 & 352.13 & 352.8 & NWS & L & 1.27 \\ 

 $2^{M,T}$   & 2010-04-05 & 95.35 & 95.52 & 96.57 & NWS & L & 0.4\\

 $3^{P}$  & 2010-05-28 & 148.11 & 148.85 & 149.7 & WSE & H & 1.01\\
 
 $4^{T,N}$   & 2010-06-20 & 171.95 & 172.35 & 173.7 & NES & L & 1.76\\
 
 $5^{T,N}$   & 2011-02-04 & 35.2  & 35.58 & 35.82 & NES & L & 0.44\\ 

 $6^{P,M,N}$   & 2011-03-29 & 88.4 & 89.01 & 91.4 & NES & L &  2.36\\

 $7^{N}$   & 2011-05-28 & 147.69 & 148.28 & 148.88 & SWN & L & 1.04\\
 
 $8^{P,T}$   & 2011-06-05 & 155.85 & 156.05 & 156.8 & WNE & H & 0.97\\
 
 $9^{P,M,T}$   & 2011-09-17 & 260.15 & 260.69 & 261.49 & SEN & L &  0.89\\ 

 $10^{P,T}$   & 2012-01-22 & 22.25 & 22.52 & 22.77 & NWS & L & 0.87\\

 $11^{P,M}$   & 2012-05-16 &  137.5 & 137.75 & 138.75 & SWN & L & 0.57\\
 
 $12^{P,M,T}$   & 2012-06-16 & 168.5 & 169.05 & 169.51  & NES & L &  1.12\\
 
 $13^{P,T,N}$   & 2012-10-08 & 282.22 & 282.8 & 283.35 & ESW & H &  1.83\\
 
 $14^{P}$   & 2012-10-12 & 286.4 & 286.7  & 387.43 & WSE & H & 1.13\\ 

 $15^{T}$   & 2013-11-12 & 317.95 & 318.4 & 319.15 & NES & L  & 0.66\\

 $16^{P}$   & 2013-01-17 & 17 & 17.71 & 15.8 & SWN & L & 0.58\\
 
 $17^{P,M,T}$  & 2013-04-13 & 103.95 & 104.75 & 105.8 & ENW & H & 2.5 \\
 
 $18^{N}$  & 2013-06-06&  157.1  & 157.96 & 159 & WSE  & H & 1.08\\
 
 $19^{P,M,T}$   & 2013-07-12 & 193.65 & 194.25 & 195.35 & NWS & L & 1.1\\ 

 $20^{M,T}$   & 2013-10-02 & 275.07 & 275.96 & 276.95 & ENW & H  & 0.78\\

 $21^{P,T}$   & 2014-08-19 & 231.3 & 231.85 & 233 & WNE & H & 0.96\\
 
 $22^{N}$ & 2016-10-12  & 286.92 & 287.25 & 288.62 & SEN & L &  1.21\\
 
\hline                  
\end{tabular}
\end{table*}
%
%-------------------------------------------------------------

The calculated NRF ratio $\theta$ for each event is given in the last column of Table \ref{tab4}. 
Figure \ref{Fig6} shows the relative number of events (i.e., occurrence frequency) separately for high (orange) and low (blue) inclination events, with respect to the calculated $\theta$ ratios. We can see that the majority of events have a $\theta$ ratio close to 1, regardless of inclination. However, we can also see that the frequency for small $\theta$ ratios ($\theta<0.75$) is higher for low inclination events, more precisely there is no high inclination event with a $\theta$ ratio smaller than 0.75. Also, the frequency for high $\theta$ ratios ($\theta>1.25$) is higher for high inclination events. This indicates that NRFs in the sheath region are more pronounced in the $\pm$ y direction for high inclination events and that NRFs are more pronounced in the $\pm$ z direction for low inclination events.
We calculated the mean value, the standard deviation, %and median of $\theta$ ratios separately for low and high inclined events. These results are shown in table \ref{tab4} and we can see that the calculated mean value and median are slightly higher for high inclination events. 
the median, and the 95\% percentiles of the $\theta$ ratios separately for low and high inclined events. The results are presented in table \ref{tab4}. We can see that the calculated mean and median are slightly higher for high inclination events; however, we cannot confirm the statistical significance due to the very low number of high inclination events.

   \begin{table}
      \caption[]{Statistical results (mean value, median, and standard deviation) derived separately for high inclination and low inclination event samples.}
         \label{tab4}
     $$ 
         \begin{array}{p{0.5\linewidth}l}
            \hline
            \noalign{\smallskip}
            \hspace{2.5cm} High      &  Low  \\
            \noalign{\smallskip}
            \hline
            \noalign{\smallskip}
            Mean \hspace{1.8cm} 1.3 & 1.0      \\
            Std \hspace{2.1cm} 0.6    & 0.5  \\
            Median \hspace{1.5cm} 1.1 & 1.0              \\
            95\% percentile \hspace{0.5cm} 2.5 & 2.2              \\
            \noalign{\smallskip}
            \hline
         \end{array}
     $$ 
   \end{table}

Nevertheless, there is an indication that in the sheath region NRFs might be more pronounced in the $\pm$ y direction for high inclination events, whereas for low inclination events they are more pronounced in the $\pm$ z direction. This asymmetry could have an implication for the CME propagation. Namely the difference in NRF flows indicates a difference in the pileup and draping of the IMF for differently oriented CMEs, which is directly related to the MHD drag. This concept is presented in Figure \ref{fig7} and is based on the previous work by \cite{Gosling1987}.

\cite{Gosling1987} argued that the IMF draping around CMEs should depend on the CME size and shape and that it can result in the enhancement of the out-of-ecliptic component ($B_z$) at the expense of the ecliptic components ($B_x$ and $B_y$). To visualize the complex 3D draping of the IMF, they considered the IMF draping in the ecliptic and out-of-ecliptic (meridional) planes separately using simplified IMF configurations of a spiral and dipole (i.e., purely radial IMF), respectively (see top panels of Figure \ref{fig7}). We expand on this interpretation by including the CME orientation. Assuming that the CME geometry can be represented as that of a toroidal FR, the shape and size of the CME front is expected to be different in the ecliptic and meridional planes and also depend on the FR orientation. This is shown in the middle and bottom panels of Figure \ref{fig7}.

In each panel of figure \ref{fig7}, the $yx$ plane represents the meridional plane, whereas the $xz$ plane represents the ecliptic plane. The viewing plane is marked in the upper right corner of each panel. The top panels in Figure \ref{fig7} show the idealized configuration of the IMF (red arrows) in the ecliptic (left) and meridional (right) plane. The middle panels show a low inclination CME embedded in an idealized IMF in the ecliptic (left) and meridional (right) plane. The bottom panels show a high inclination CME embedded in an idealized IMF, again, in the ecliptic (left) and meridional (right) plane. The blue arrows in the middle and bottom panels mark the direction of the deflection of the ambient plasma away from the path of the CME in the east-west and north-south direction for ecliptic and meridional planes, respectively.

Fast CMEs interact with the ambient solar wind plasma and the IMF as they propagate. A slower moving ambient plasma ahead of the CME is accelerated and deflected from its path. Assuming a toroidal shape for the FR, its leading surface is characterized by two curvatures: 1) an axial curvature due to the rooting of the footpoints of the CME at the Sun; and 2) a cross-sectional curvature due to its internal magnetic field
structure. Axial curvature is greater in extent which means it has smaller curvature radii in comparison to the cross-sectional curvature which is smaller in extent and has greater curvature radii. The ambient plasma is expected to be more easily deflected via cross-sectional curvature due to its smaller extent (greater curvature radii) in comparison to the axial curvature. For better understanding, we can draw an analogy with a ship on the water. Namely, the front part of every ship is very small in extent, and this is to allow the water in front of the ship to flow more easily around it.

Under the assumption that the IMF is “frozen” in the ambient solar wind, a draping of the IMF occurs. IMF draping around the traveling transients in the heliosphere, such as CMEs, is essentially a consequence of the fact that magnetized plasma cannot significantly penetrate into the transient, and thus it is forced to flow around it \citep{Gosling1987}. As a result, for a low inclination CME, we might expect the ambient plasma to be more easily deflected in the north-south direction where the CME extent is smaller, that is we might expect $\theta<1$. The more complex spiral-structured IMF can thus, following the deflected plasma, more easily escape via the meridional extent of the CME and thus be out of the CME's path. Consequently, there would be less draping of the more complex  spiral-structured IMF across the CME front. For high inclined CMEs, the situation is the reverse as they have a much wider spread in the meridional plane compared to the ecliptic plane. Here we might expect the ambient plasma to be more easily deflected in the east-west direction where the CME size is smaller, that is we might expect $\theta>1$. The more complex spiral-structured IMF cannot easily escape via the meridional extent of the CME and thus it is out of the CME's path. Consequently, there would be more draping of the more complex  spiral-structured IMF across the CME front. Colloquially put, the CME should be able to ``swim'' more easily when it has a low inclination.
 
 However, as shown in  \cite{Schwenn2006}, the velocity of the ambient solar wind is lower near the ecliptic plane than in the higher latitude regions. Consequently, when considering fast CMEs, the relative velocity of a CME and ambient solar wind is larger in low latitude regions. Since the drag force increases as the relative velocity increases (\citealt{Cargill2004};  \citealt{Vrsnak2001}; \citealt{Chen1993}), it is to be excepted that low inclination events experience greater drag force due to a much wider spread in the ecliptic plane than high inclination events.

\section{Summary and conclusions}

We analyzed 22 well-associated CME-ICME pairs during the rising and the maximum phase of solar cycle 24. We determined their inclination, both at Sun and in situ at the Lagrange L1 point. We derived the CME tilt close to the Sun using three different techniques: GCS, C2-ellipse fitting, and C3-ellipse fitting. GCS was performed when images from at least two spacecraft were available for the 3D reconstruction, while the ellipse fit was performed using the single LASCO-C2 and LASCO-C3 coronagraphic images. The in situ FR type was determined by visual inspection of the magnetic field components in the GSE coordinate system. 

Comparing our GCS and ellipse fit results for the FR inclination with results from the \href{https://www.helcats-fp7.eu/catalogues/wp3_kincat.html}{HELCATS catalog}, \cite{Temmer2021}, and \cite{Sachdeva2019}, we concluded that the methods are only robust enough to determine whether the FR is of a dominantly high or low inclination. In accordance with this, we only distinguished low and high incline events from the in situ data. 
When comparing the results for high and low inclination at the near-Sun and at the near-Earth environment, we found that the majority of the events, $68\%$,
have a consistent estimation for the tilt from remote and in situ data. 
Also, we found that the majority, $73\%$, have a low inclination. 
We showed that the CMEs' tilt obtained by GCS varies greatly when determined by different observers, as well as that GCS results are different from the results obtained by the C2- and C3-ellipse fitting technique. These results show that the CMEs' tilt determination still remains a challenge. 

Our analysis of the NRFs in the sheath region indicates that high inclination events (as observed in situ) show a slightly higher velocity ratio of the y to z direction. This suggests that the NRFs in the sheath region of high inclined CMEs are more profound in the east-west direction than in the case of low inclined events. Thus, for low inclined events, we might expect the more complex spiral-structured IMF to more easily escape via the meridional extent of the CME and thus be out of the CME's path. This result shows the potential for further research on the relation between an ICME's inclination and propagation. However, in order to do so, much larger sample sizes are needed to provide results of high statistical significance. Due to all the restrictions imposed by different tilt determination methods, sample size increment is certainly not an easily achievable task. However, we have shown here that the C2- and C3-ellipse fit techniques provide results for inclination in good agreement with 3D CME reconstruction using GCS. Thus, the sample sizes can be significantly increased in future work by analyzing CME-ICME pairs before the STEREO era.

\begin{acknowledgements}

We acknowledge the support by the Croatian Science Foundation under the project IP-2020-02-9893(ICOHOSS). K.M. acknowledges support by Croatian Science Foundation in the scope of Young Researches Career Development Project Training New Doctoral Students.

The SOHO/LASCO data used here are produced by a consortium of the Naval Research Laboratory (USA), Max-Planck-Institut fuer Aeronomie (Germany), Laboratoire d’Astronomie (France), and the University of Birmingham (UK). SOHO is a project of international cooperation between ESA and NASA.

We acknowledge the STEREO/SECCHI consortium for providing the data. The SECCHI data used here were produced by an international consortium of the Naval Research Laboratory (USA), Lockheed Martin Solar and Astrophysics Lab (USA), NASA Goddard Space Flight Center (USA), Rutherford Apple- ton Laboratory (UK), University of Birmingham (UK), Max- Planck-Institute for Solar System Research (Germany), Centre Spatiale de Liege (Belgium), Institut d’Optique Theorique et Appliquee (France), Institut d’Astrophysique Spatiale (France).

We acknowledge use of NASA/GSFC’s Space Physics Dana Facility’s OMNIWeb (or CDAWeb or ftp) service, and OMNI data.

B.V. also acknowledges support by the Croatian Science Foundation under the project 7549 ”Millimeter and sub-millimeter observations of the solar chromosphere with ALMA

\end{acknowledgements}

   \bibliographystyle{aa} % style aa.bst
   \bibliography{43433corr} % your references Yourfile.bib

%\begin{thebibliography}{}
%
%  \bibitem[Baker(1966)]{baker} Baker, N. 1966,
%      in Stellar Evolution,
%      ed.\ R. F. Stein,\& A. G. W. Cameron
%      (Plenum, New York) 333
%
%   \bibitem[Balluch(1988)]{balluch} Balluch, M. 1988,
%      A\&A, 200, 58
%
%   \bibitem[Cox(1980)]{cox} Cox, J. P. 1980,
%      Theory of Stellar Pulsation
%      (Princeton University Press, Princeton) 165
%
%   \bibitem[Cox(1969)]{cox69} Cox, A. N.,\& Stewart, J. N. 1969,
%      Academia Nauk, Scientific Information 15, 1
%
%   \bibitem[Mizuno(1980)]{mizuno} Mizuno H. 1980,
%      Prog. Theor. Phys., 64, 544
%   
%   \bibitem[Tscharnuter(1987)]{tscharnuter} Tscharnuter W. M. 1987,
%      A\&A, 188, 55
%  
%   \bibitem[Terlevich(1992)]{terlevich} Terlevich, R. 1992, in ASP Conf. %Ser. 31, 
%      Relationships between Active Galactic Nuclei and Starburst %Galaxies, 
%      ed. A. V. Filippenko, 13
%
%   \bibitem[Yorke(1980a)]{yorke80a} Yorke, H. W. 1980a,
%      A\&A, 86, 286
%
%   \bibitem[Zheng(1997)]{zheng} Zheng, W., Davidsen, A. F., Tytler, D. \& %Kriss, G. A.
%      1997, preprint
%\end{thebibliography}

\end{document}